\documentclass{PoS}

\usepackage{subfigure}

\newcommand{\be}{\begin{eqnarray}}
\newcommand{\ee}{\end{eqnarray}}
\newcommand{\nee}{\nonumber\end{eqnarray}}
\newcommand{\nn}{\nonumber\\}

\newcommand{\drbar}{{\overline{\rm DR}}}

\newcommand{\mch}[1] {m_{\ti \x^+_{#1}}}
\newcommand{\mnt}[1] {m_{\ti \x^0_{#1}}}

\newcommand{\msg}    {m_{\ti g}}
\newcommand{\msu}[1] {m_{\ti u_{#1}}}
\newcommand{\msd}[1] {m_{\ti d_{#1}}}

\def\gev             {{\rm GeV}}

\def\be            {\begin{equation}}
\def\ee            {\end{equation}}
\def\bea            {\begin{eqnarray}}
\def\eea            {\end{eqnarray}}

\def\d               {\delta}

\def\x               {\chi}
\def\ti              {\tilde}
\def\sq              {\ti q}
\def\st              {\ti t}
\def\sb              {\ti b}

\def\sg              {\ti g}

\def\su                {\ti{u}}

\def \sca                 {\ti{c}}
\def\sd                {\ti{d}}
\def\ss                  {\ti{s}}

\def\dll            {\d^{LL}_{23}}
\def\durr            {\d^{uRR}_{23}}
\def\durl            {\d^{uRL}_{23}}
\def\dulr            {\d^{uLR}_{23}}
\def\ddrr            {\d^{dRR}_{23}}
\def\ddrl            {\d^{dRL}_{23}}
\def\ddlr            {\d^{dLR}_{23}}
\title{Gluino two-body decays at full one-loop level in the MSSM with quark-flavour violation}

\ShortTitle{Gluino two-body decays at full one-loop level in the MSSM with QFV}

\author{\speaker{Helmut Eberl}, Elena Ginina,\\
       Institut f\"ur Hochenergiephysik der \"O{}sterreichischen Akademie der Wissenschaften\\
       E-mail: \email{helmut.eberl@oeaw.ac.at},  \email{elena.ginina@oeaw.ac.at}}
       
\author{Keisho Hidaka\\
        Department of Physics, Tokyo Gakugei University, Koganei, Tokyo 184-8501, Japan\\
        E-mail: \email{hidaka@u-gakugei.ac.jp}}

\abstract{We study the two-body decays of the gluino at full one-loop level in the Minimal Supersymmetric Standard Model with quark-flavour violation (QFV) in the squark sector. The renormalization is done in the $\drbar$ scheme and hard gluon and photon radiations are included by adding the corresponding three-body decay widths. In the numerics the dependence of the gluino decay widths on the QFV parameters is discussed
with special emphasis put on the separation of the electroweak and the SUSY QCD corrections. 
The main dependence stems from the $\tilde c_R - \tilde t_R$ mixing in the decays to up-type squarks  
because there hold strong constraints from B-physics on the other quark-flavour mixing parameters. 
Including the full one-loop corrections, the changes of the gluino decay widths are mostly negative  and of the order of about -10\%.
The QFV part stays small in the total width but can vary up to -8\% for the decay width into the lightest squark.
For the corresponding branching ratio, however, the effect is smaller by at least a factor of two. 
The electroweak corrections can become 35\% of the SUSY QCD corrections.}

\FullConference{13th International Symposium on Radiative Corrections (Applications of Quantum Field Theory to Phenomenology)\\
         25-29 September, 2017 \\
         St. Gilgen, Austria}

\begin{document}

\section{Introduction}
Although there is no sign of new particles yet, the MSSM is still favoured as a discoverable theory beyond the SM and 
will be searched for with high priority at all experiments. SUSY particle decay chains have been extensively studied during the last two decades.
Especially relevant are the decays of strongly interacting squarks and gluinos because they can be produced at LHC copiously.
There exist an enormous number of MSSM studies. Nevertheless it has yet unstudied potential related to more general 
treatment of its squark sector parameters. Despite the stringent constraints from B and K physics, such parameters can lead to quark-flavour violation (QFV) and can change the phenomenological observables significantly. 

In~\cite{Beenakker:1996dw} the two-body decays of the gluino in the MSSM in the quark flavour conserving (QFC) 
case were studied including the one-loop SUSY-QCD
corrrections, and the full one-loop corrections allowing complex parameters were presented in~\cite{Heinemeyer:2011ab}.
Studies of these decays including general quark-flavour violation in the squark sector at tree-level are in~\cite{Hurth:2009ke,Bartl:2009au}.
In~\cite{Bartl:2011wq} three-body decays of the gluino at tree-level are discussed.

In this work based on~\cite{sg-paper} we study the two-body decays of the gluino at full one-loop level in the MSSM with quark-flavour violation in the squark sector.
For further details see also the PhD thesis of Sebastian Frank~\cite{SebastianPhDth}.
The renormalization is done in the $\drbar$ scheme and hard gluon and photon radiations are included by adding the corresponding three-body decay widths.

\section{QFV parameters}
\noindent
In the SM QFV is within the Cabibbo-Kobayashi-Maskawa (CKM) matrix. 
In the general MSSM there are two concepts:
\begin{enumerate}
\item {\it Minimal quark flavour violation} - no new sources of QFV, in the super-CKM basis the squarks undergo the same rotations like the quarks, 
all flavour violating entries are related to the CKM matrix,
\item  {\it Non-minimal quark flavour violation} - new sources of QFV, independent of the CKM, considered as free parameters in the theory.
\end{enumerate}
In this study 2. is assumed.\\

\noindent
The hermitian $6\times6$ squark mass matrix in the super-CKM basis is
\begin{equation}
    {\cal M}^2_{\tilde{q}} = \left( \begin{array}{cc}
        {\cal M}^2_{\tilde{q},LL} & {\cal M}^2_{\tilde{q},LR} \\[2mm]
        {\cal M}^2_{\tilde{q},RL} & {\cal M}^2_{\tilde{q},RR} \end{array} \right)\,,
 \label{EqMassMatrix1}
\end{equation}
with $\sq = \su, \sd$. The $3 \times 3$ blocks in eq.~(\ref{EqMassMatrix1}) for the up squarks are
\begin{eqnarray}
    & &{\cal M}^2_{\tilde{u},LL} = V_{\rm CKM} M_Q^2 V_{\rm CKM}^{\dag} + D_{\tilde{u},LL}{\bf 1} + \hat{m}^2_u, \nonumber \\
    & &{\cal M}^2_{\tilde{u},RR} = M_U^2 + D_{\tilde{u},RR}{\bf 1} + \hat{m}^2_u, \nonumber \\
    & & {\cal M}^2_{\tilde{u},RL} = {\cal M}^{2\dag}_{\tilde{u},LR} =
    \frac{v_2}{\sqrt{2}} T_U - \mu^* \hat{m}_u\cot\beta\, .
     \label{RLblocks}
\end{eqnarray}
$v_{2}=\sqrt{2} \left\langle H^0_{2} \right\rangle$, 
$M^2_{Q,U}$ are hermitian soft SUSY-breaking squark mass matrices and $T_{U}$ is a soft SUSY-breaking trilinear 
coupling matrix. 
The ratio of the vacuum expectation values of the neutral Higgs fields $v_2/v_1 = \tan\beta$, $\mu$ is the higgsino mass parameter, 
and $\hat{m}_{u,d}$ are the diagonal mass matrices of the up- and down-type quarks. 
Furthermore, 
$D_{\tilde{q},LL} = \cos 2\beta m_Z^2 (T_3^q-e_q
\sin^2\theta_W)$ and $D_{\tilde{q},RR} = e_q \sin^2\theta_W \cos 2\beta m_Z^2$,
where
$T_3^q$ and $e_q$ are the isospin and
electric charge of the quarks (squarks), respectively, and $\theta_W$ is the weak mixing
angle. We approximate the $V_{\rm CKM}$ by the unit matrix.
Analogous formulas hold for the down squarks.\\
The eigenvalue problems for up and down squarks can be written as
\begin{equation}
\begin{array}{c}
U^{\tilde{u}} {\cal M}^2_{\tilde{u}} (U^{\tilde{u} })^{\dag} = {\rm diag}(m_{\tilde{u}_1}^2,\dots,m_{\tilde{u}_6}^2)\, , \\[3mm]
U^{\tilde{d}} {\cal M}^2_{\tilde{d}} (U^{\tilde{d} })^{\dag} = {\rm diag}(m_{\tilde{d}_1}^2,\dots,m_{\tilde{d}_6}^2) \, , 
\end{array}
\, {\rm with} \,
\left(
\begin{array}{c}
\tilde u_1 \\ \tilde u_2 \\ \tilde u_3 \\ \tilde u_4 \\ \tilde u_5 \\ \tilde u_6
\end{array}
\right) = U^{\tilde u} .
\left(
\begin{array}{c}
\tilde u_L \\ \tilde c_L \\ \tilde t_L \\ \tilde u_R \\ \tilde c_R \\ \tilde t_R
\end{array}
\right)\, , \,
\left(
\begin{array}{c}
\tilde d_1 \\ \tilde d_2 \\ \tilde d_3 \\ \tilde d_4 \\ \tilde d_5 \\ \tilde d_6
\end{array}
\right) = U^{\tilde d} .
\left(
\begin{array}{c}
\tilde d_L \\ \tilde s_L \\ \tilde b_L \\ \tilde d_R \\ \tilde s_R \\ \tilde b_R
\end{array}
\right)\, .
\end{equation}
QFV is expressed by dimensionless parameters, 
in the up-type squark sector they are
\begin{eqnarray}
\delta^{LL}_{\alpha\beta} & \equiv & M^2_{Q \alpha\beta} / \sqrt{M^2_{Q \alpha\alpha} M^2_{Q \beta\beta}}~,
\label{eq:InsLL}\\[3mm]
\delta^{uRR}_{\alpha\beta} &\equiv& M^2_{U \alpha\beta} / \sqrt{M^2_{U \alpha\alpha} M^2_{U \beta\beta}}~,
\label{eq:InsRR}\\[3mm]
\delta^{uRL}_{\alpha\beta} &\equiv& (v_2/\sqrt{2} ) T_{U\alpha \beta} / \sqrt{M^2_{U \alpha\alpha} M^2_{Q \beta\beta}}~,
\label{eq:InsRL}
\end{eqnarray}
with $\alpha,\beta=1,2,3 ~(\alpha \ne \beta)$ denoting the quark flavours $u,c,t$.
Analogous formulas hold for the sdown parameters $\delta^{dRR}_{\alpha\beta}$ and $\delta^{dRL}_{\alpha\beta}$.

\section{Gluino two-body decay widths}
\label{sec:process}
\noindent
The explicit form of the tree-level width is
\bea
\Gamma^0(\sg \to \sq^*_i q_g) & = & {\lambda^{1/2}(\msg^2,m_{\sq_i}^2, m_{q_g}^2) \over 32\,  \msg^3} \alpha_s\, \bigg(
\left(|U^{\sq}_{i,g}|^2+|U^{\sq}_{i,g+3}|^2\right)(\msg^2-m_{\sq_i}^2 +m_{q_g}^2) \\
&& \hspace*{7cm}- 4 \msg m_{q_g}{\rm Re}\left(U^{\sq *}_{i,g}\, U^{\sq}_{i,g+3} \right) \bigg) \nonumber
\label{decaywidttree}
\eea
with $i=1,...,6$ (no summation), $q = u, d$, and the subscript $g$ is the quark-generation index.

In order to obtain an ultraviolet (UV) convergent result at one-loop level we employ the SUSY invariant 
dimensional reduction ($\overline {\rm DR}$) regularisation scheme, which implies that all tree-level input parameters of the Lagrangian are defined at the scale $Q=M_3\approx m_{\sg}$. Since in this scheme the tree-level couplings 
$g^i_{L,R}$ are defined at the scale $Q$, they do not receive further finite shifts due to radiative corrections. The physical scale independent masses and fields are obtained from the $\overline {\rm DR}$ ones using on-shell renormalisation conditions. \\

\begin{figure*}[h!]
{\centering \mbox{\hspace*{-1cm} \resizebox{17cm}{!}{\includegraphics{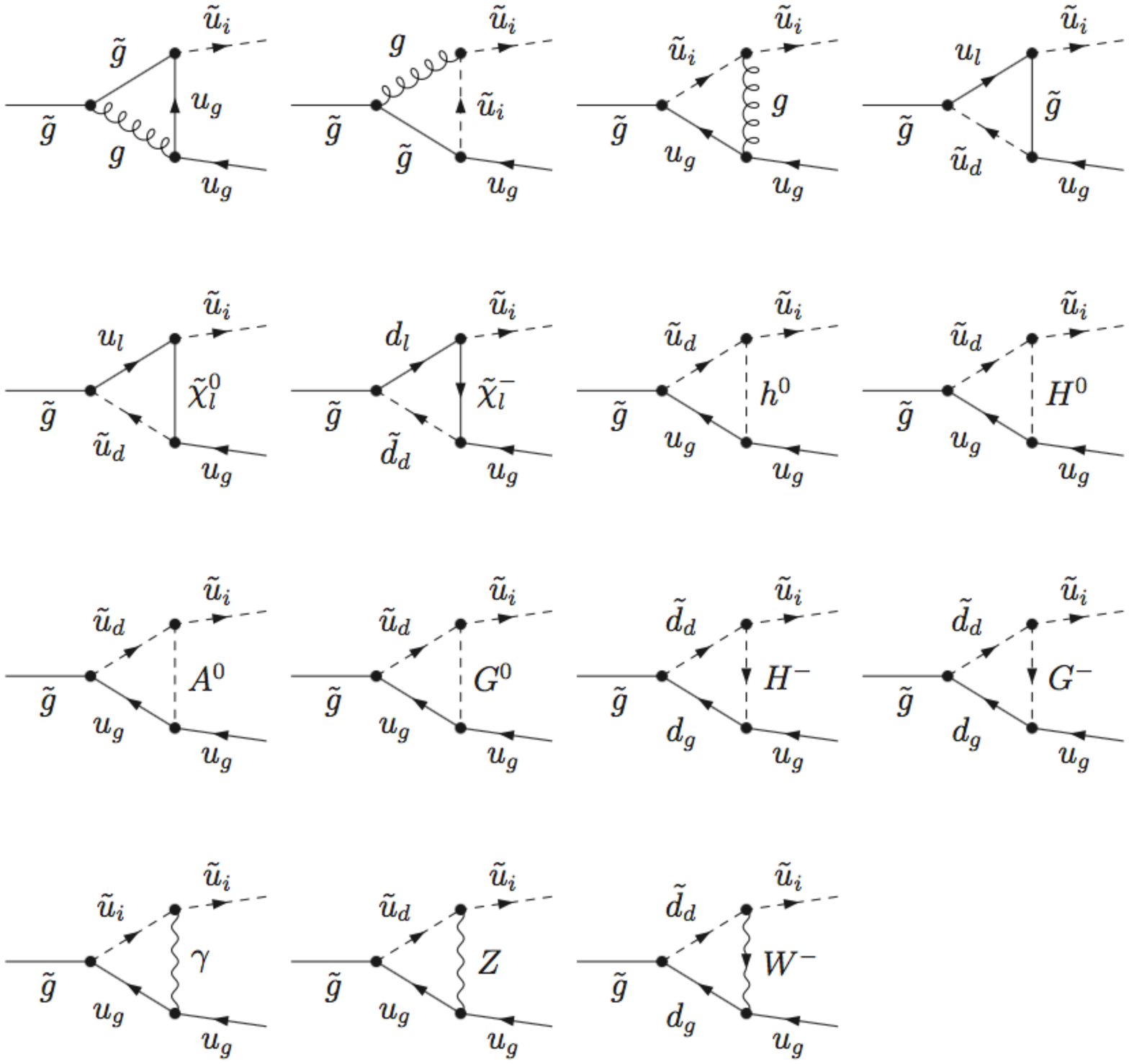}} }
   \label{fig1a}}
\caption{Feynman graphs of all one-loop vertex contributions to $\Gamma(\sg \to \sq^*_i q_g),\, i = 1,\ldots, 6,\, g = 1,2,3$.}
\label{vertex-corr}
\end{figure*}

\noindent
The renormalised one-loop partial decay widths are given by
\bea
\label{decaywidth}
\hspace*{-2cm}
\Gamma(\sg \to \sq^*_i q_g) &=& \Gamma^0( \sg \to \sq^*_i q_g)~+~\Delta \Gamma(\sg \to \sq^*_i q_g)\, , \quad {\rm with}  \\
\Delta \Gamma(\sg \to \sq^*_i q_g)  &= & {\lambda^{1/2}(\msg^2,m_{\sq_i}^2, m_{q_g}^2) \over 512\, \pi\,  \msg^3} {\rm Re}({\cal M}_0^\dagger {\cal M}_1)\,,
\quad {\rm and} \nn
 {\rm Re}({\cal M}_0^\dagger {\cal M}_1) & = & {\rm Re}\bigg( (g^{i*}_L \Delta g_L + g^{i*}_R \Delta g_R)(\msg^2-m_{\sq_i}^2 +m_{q_g}^2) + 2 \msg m_{q_g}(g^{i*}_L \Delta g_R +  g^{i*}_R \Delta g_L)\bigg) \nonumber\, ,
\eea
where $ {\cal M}_0$ is the tree-level, and $ {\cal M}_1$ the one-loop amplitude. 
The $Q$ independent renormalised shifts to the left and right couplings can be split,  
\be
\Delta g_{L,R} = \d g_{L,R}^v + \d g_{L,R}^w +\d g_{L,R}^c\,,
\ee
with $v$ stands for the vertex corrections, $w$ for the corrections of the wave functions, and $c$ for the coupling counter terms.
Futher details including also the treatment of the infrared divergences can be found in~\cite{sg-paper} and in ~\cite{SebastianPhDth}.
The full one-loop two-body decay width eq.~(\ref{decaywidth}) includes one-loop shifts from SQCD (gluon and gluino) and EW (electroweak including also the photon) corrections,
\be
\Gamma(\sg \to \sq^*_i q_g) = \Gamma^0(\sg \to \sq^*_i q_g)+
\Delta \Gamma^{\rm SQCD}(\sg \to \sq^*_i q_g) + \Delta \Gamma^{\rm EW}(\sg \to \sq^*_i q_g)\,.
\ee
For illustration, all one-loop vertex graphs are shown in Fig~\ref{vertex-corr}. The first line gives contributions to SQCD and the other ones to EW.

In the numerical study the notation $\Gamma(\sg \to \sq_i q_g) = \Gamma(\sg \to \sq^*_i q_g) + \Gamma(\sg \to \sq_i \bar q_g)$ 
is used. This is equivalent with $2 \Gamma(\sg \to \sq^*_i q_g)$ when CP is conserved, which is the case in this study.

\section{Numerical results}
The gluino two-body widths and branching ratios are calculated with the publicly available numerical code FVSFOLD, developed by Sebastian Frank~\cite{SebastianPhDth}. For building FVSFOLD also the packages FeynArts~\cite{Hahn:2000kx,Hahn2} and 
FormCalc~\cite{Hahn:1998yk} were used. FVSFOLD needs LoopTools~\cite{Hahn:1998yk} based on FF~\cite{FF}.
The counter terms are included by ourselves. 
The code used here is cross-checked by SARAH (personal correspondence with F.~Staub).
For calculating the $h^0$ mass and the low-energy observables, especially those ones in 
the B-sector we use \mbox{SPheno v3.3.3}~\cite{SPheno1, SPheno2}. The experimental and theoretical constraints taken into account are 
presented in Appendix~B of~\cite{sg-paper}.\\

First we fix a reference scenario which fulfills all constraints, see Table~\ref{basicparam}. 
Then we vary the QFV parameters within the allowed region. The MSSM mass spectrum is presented in Table~\ref{physmasses}. We see that our 
reference point violates QF explicitly and we have approximately gauge unification. 
Table~\ref{flavourdecomp} shows the flavour decompositions of the $\su_{1,2}$ and $\sd_{1,2}$ squarks. The $\su_1$ squark is a strong mixture of 
$\sca_R$ and $\st_R$, with a tiny contribution from $\sca_L$, and the $\su_2$ squark is mainly $\st_L$, with a tiny contribution from $\sca_R$. 
The $\sd_1$ is a mixture of $\ss_R$ and $\sb_R$, and $\sd_2$ is a pure  $\sb_L$. The decays of $\sg$ into the lightest two sups and sdowns
are possible with the branching ratios B$(\sg \to \su_1\,c)\approx 17\%$, B$(\sg \to \sd_1\,s)\approx 18\%$, B$(\sg \to \su_1\,t) = {\rm B}(\sg \to \sd_1\,b) \approx 27\%$, B$(\sg \to \su_2\,t)\approx 5\%$. The total two-body width including the full one-loop contribution, 
$\Gamma(\sg \to \sq q) = 70\,\gev$. The tree-level width $\Gamma^0(\sg \to \sq q) = 75\,\gev$.\\

The QFV parameters 
$\dulr, \durl, \ddlr$, and $\ddrl$ are constrained from the 
vacuum stability conditions. Thus they must remain rather small. 
A large $\dll$ is not possible because it violates 
B-physics constraints such as that for B($B_s \to \mu^+ \mu^-$). 
However, large right-right mixing in both 
$\su$ and $\sd$ sectors is allowed.
Therefore, in the following we only need to show diagrams with dependences on 
$\durr$ and $\ddrr$.\\

\begin{table}[h!]
\caption{QFV reference scenario: all parameters are calculated
at $Q = M_3 = 3~{\rm TeV} \simeq m_{\sg}$,
except for $m_{A^0}$ which is the pole mass of $A^0$, 
and $T_{U33} =2500$~GeV (corresponding to $\delta^{uRL}_{33} = 0.06$). All other squark parameters are zero.}
\begin{center}
\begin{tabular}{|c|c|c|c|c|c|}
  \hline
 $M_1$ & $M_2$ & $M_3$ &$\mu$ & $\tan \beta$ & $m_{A^0}$ \\
 \hline \hline
 500~\gev  &  1000~\gev & 3000~\gev &500~\gev & 15 &  3000~\gev \\
  \hline
\end{tabular}
\vskip0.4cm
\begin{tabular}{|c|c|c|c|}
  \hline
   & $\alpha = 1$ & $\alpha= 2$ & $\alpha = 3$ \\
  \hline \hline
   $M_{Q \alpha \alpha}^2$ & $3200^2~\gev^2$ &  $3000^2~\gev^2$  & $2600^2~\gev^2$ \\
   \hline
   $M_{U \alpha \alpha}^2$ & $3200^2~\gev^2$ & $3000^2~\gev^2$ & $2600^2~\gev^2$ \\
   \hline
   $M_{D \alpha \alpha}^2$ & $3200^2~\gev^2$ & $3000^2~\gev^2$ &  $2600^2~\gev^2$  \\
   \hline
\end{tabular}
\vskip0.4cm
\begin{tabular}{|c|c|c|c|c|c|c|}
  \hline
   $\delta^{LL}_{23}$ & $\delta^{uRR}_{23}$  &  $\delta^{uRL}_{23}$ & $\delta^{uLR}_{23}$ &$\delta^{dRR}_{23}$  &  $\delta^{dRL}_{23}$ & $\delta^{dLR}_{23}$\\
  \hline \hline
    0.01 & 0.7 & 0.04  & 0.07 & 0.7 & 0  & 0   \\
    \hline
\end{tabular}
\end{center}
\label{basicparam}
\end{table}
%
\begin{table}[h!]
\caption{Physical masses of the particles in GeV for the scenario of Table~\ref{basicparam}.}
\begin{center}
\begin{tabular}{|c|c|c|c|c|c|}
  \hline
  $\mnt{1}$ & $\mnt{2}$ & $\mnt{3}$ & $\mnt{4}$ & $\mch{1}$ & $\mch{2}$ \\
  \hline \hline
  $460$ & $500$ & $526$ & $1049$ & $493$ & $1049$ \\
  \hline
\end{tabular}
\vskip 0.4cm
\begin{tabular}{|c|c|c|c|c|}
  \hline
  $m_{h^0}$ & $m_{H^0}$ & $m_{A^0}$ & $m_{H^+}$ \\
  \hline \hline
  $125$  & $3000$ & $3000$ & $3001$ \\
  \hline
\end{tabular}
\vskip 0.4cm
\begin{tabular}{|c|c|c|c|c|c|c|}
  \hline
  $\msg$ & $\msu{1}$ & $\msu{2}$ & $\msu{3}$ & $\msu{4}$ & $\msu{5}$ & $\msu{6}$ \\
  \hline \hline
  $3154$ & $1602$ & $2686$ & $3087$ & $3295$ & $3300$ & $3692$ \\
  \hline
\end{tabular}
\vskip 0.4cm
\begin{tabular}{|c|c|c|c|c|c|}
  \hline
   $\msd{1}$ & $\msd{2}$ & $\msd{3}$ & $\msd{4}$ & $\msd{5}$ & $\msd{6}$ \\
  \hline \hline
  $1662$ & $2689$ & $3087$ & $3295$ & $3301$ & $3747$ \\
  \hline
\end{tabular}
\end{center}
\label{physmasses}
\end{table}
%
\begin{table}[h!]
\caption{Flavour decomposition of $\su_{1,2}$ and $\sd_{1,2}$ for the scenario of Table~\ref{basicparam}. Shown are the squared coefficients. }
\begin{center}
\begin{tabular}{|c|c|c|c|c|c|c|c|}
  \hline
  & $\su_L$ & $\sca_L$ & $\st_L$ & $\su_R$ & $\sca_R$ & $\st_R$ \\
  \hline \hline
 $\su_1$  & $0$ & $0.004$ & $0$ & $0$ & $0.38$ & $0.61$ \\
  \hline 
  $\su_2$  & $0$ & $0.001$ & $0.99$ & $0$ & $0.006$ & $0$ \\
  \hline
\end{tabular}
\vskip 0.4cm
\begin{tabular}{|c|c|c|c|c|c|c|c|}
  \hline
  & $\sd_L$ & $\ss_L$ & $\sb_L$ & $\sd_R$ & $\ss_R$ & $\sb_R$ \\
  \hline \hline
 $\sd_1$  & $0$ & $0$ & $0$ & $0$ & $0.4$ & $0.6$ \\
  \hline 
  $\sd_2$  & $0$ & $0$ & $1$ & $0$ & $0$ & $0$ \\
  \hline
\end{tabular}
\end{center}
\label{flavourdecomp}
\end{table}

\clearpage
%
\begin{figure*}[h!]
\centering
\subfigure[]{
   { \mbox{\hspace*{-1cm} \resizebox{7.5cm}{!}{\includegraphics{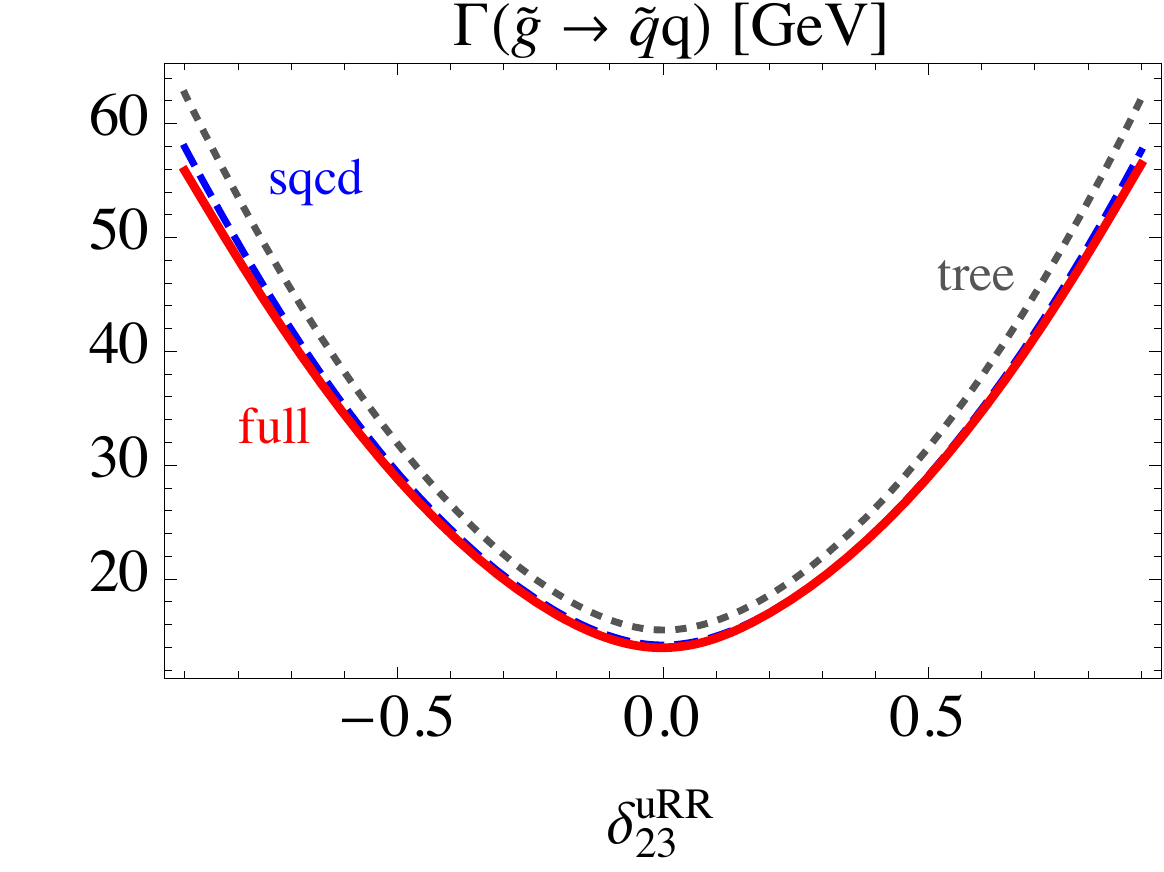}} \hspace*{-0.8cm}}}
   \label{fig1a}}
\subfigure[]{
   { \mbox{\hspace*{0.5cm} \resizebox{8cm}{!}{\includegraphics{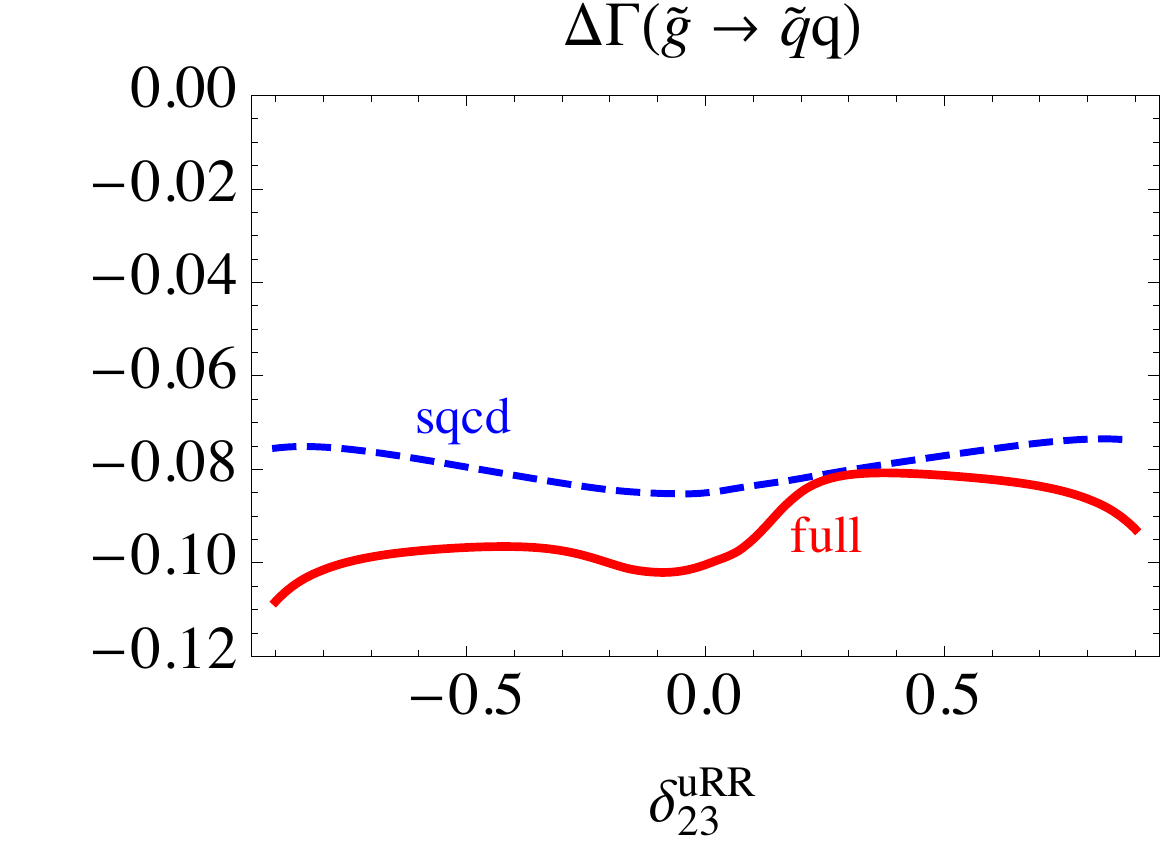}} \hspace*{-0.8cm}}}
   \label{fig1b}}\\
\subfigure[]{
   { \mbox{\hspace*{-1cm} \resizebox{7.5cm}{!}{\includegraphics{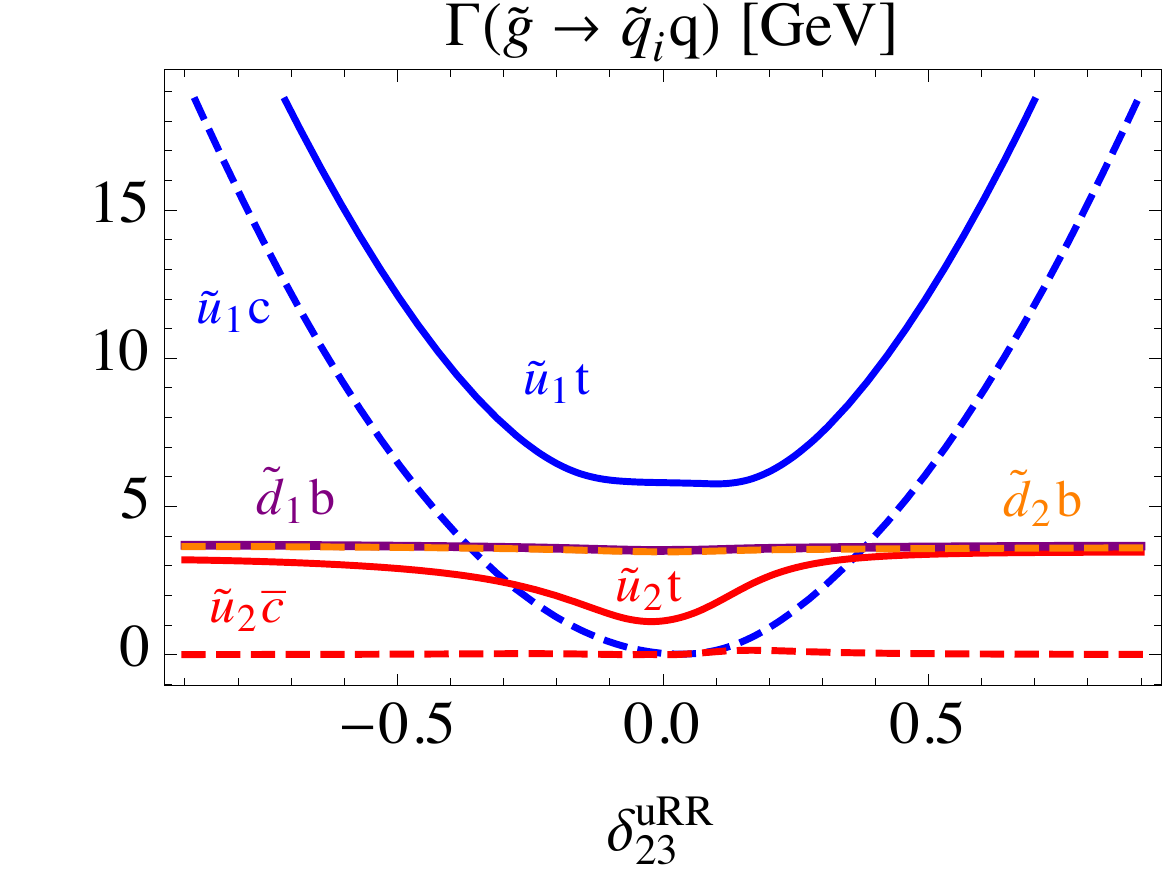}} \hspace*{-0.8cm}}}
   \label{fig1c}}
 \subfigure[]{
   { \mbox{\hspace*{+0.5cm} \resizebox{7.5cm}{!}{\includegraphics{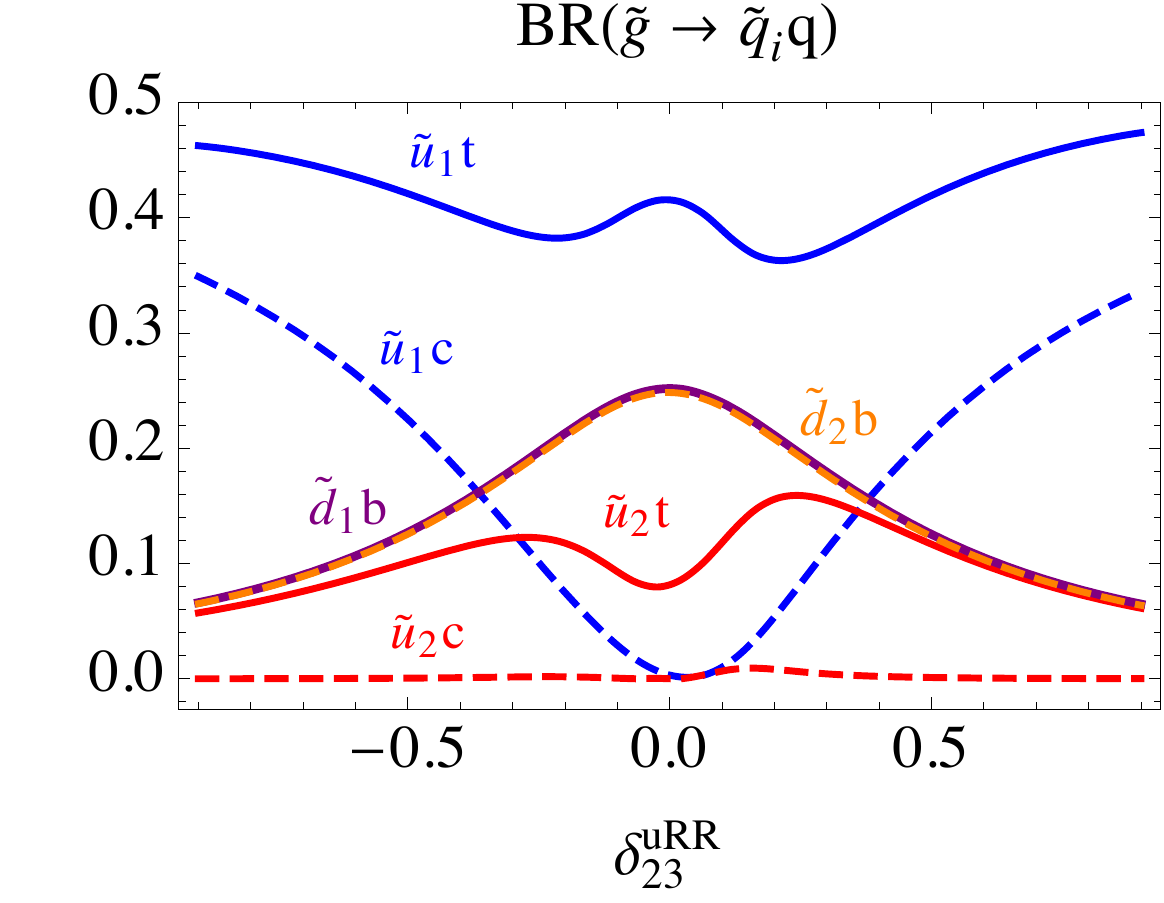}} \hspace*{-1cm}}}
  \label{fig1d}}
\caption{(a) Total two-body decay width $\Gamma(\sg \to \sq q)$
at tree-level, SQCD one-loop and full one-loop corrected as
functions of the QFV parameter $\durr$;
(b) $\Delta \Gamma(\sg \to \sq q)$ being the SQCD one-loop and the
full one-loop corrections to $\Gamma(\sg \to \sq q)$ relative to
the tree-level width; (c) Partial decay widths and (d) branching
ratios of the kinematically allowed individual two-body channels
at full one-loop level as functions of $\durr$.
All the other parameters are fixed as in Table~\ref{basicparam},
except $\durl = \dulr = 0.03$.}
\label{fig1}
\end{figure*}
We first discuss the  dependences on the QFV parameter $\durr$.
Fig.~\ref{fig1a} shows a strong dependence of $\Gamma(\sg \to \sq q)$
which stems mainly from the kinematic prefactor, see Section~\ref{sec:process}. 
The SQCD correction shown in~Fig.~\ref{fig1b} is only weakly dependent on 
$\durr$ and is about -8\%. The EW correction can become -3\% for large and negative values of $\durr$. In Fig.~\ref{fig1c} the partial widths of the 
$\sd_{1,2} b$ modes coincide because $m_{\sd_1}\approx m_{\sd_2}$.
The same holds for the branching ratios in Fig.~\ref{fig1d} . For $\durr \approx 0$ the width of $\tilde g \to \tilde u_1 c$ becomes tiny because then $\tilde u_1$ is mainly $\tilde t_R$ as all the other QFV $\delta$'s are relatively small.

%
\begin{figure*}[h!]
\centering
\subfigure[]{
   { \mbox{\hspace*{-1cm} \resizebox{8cm}{!}{\includegraphics{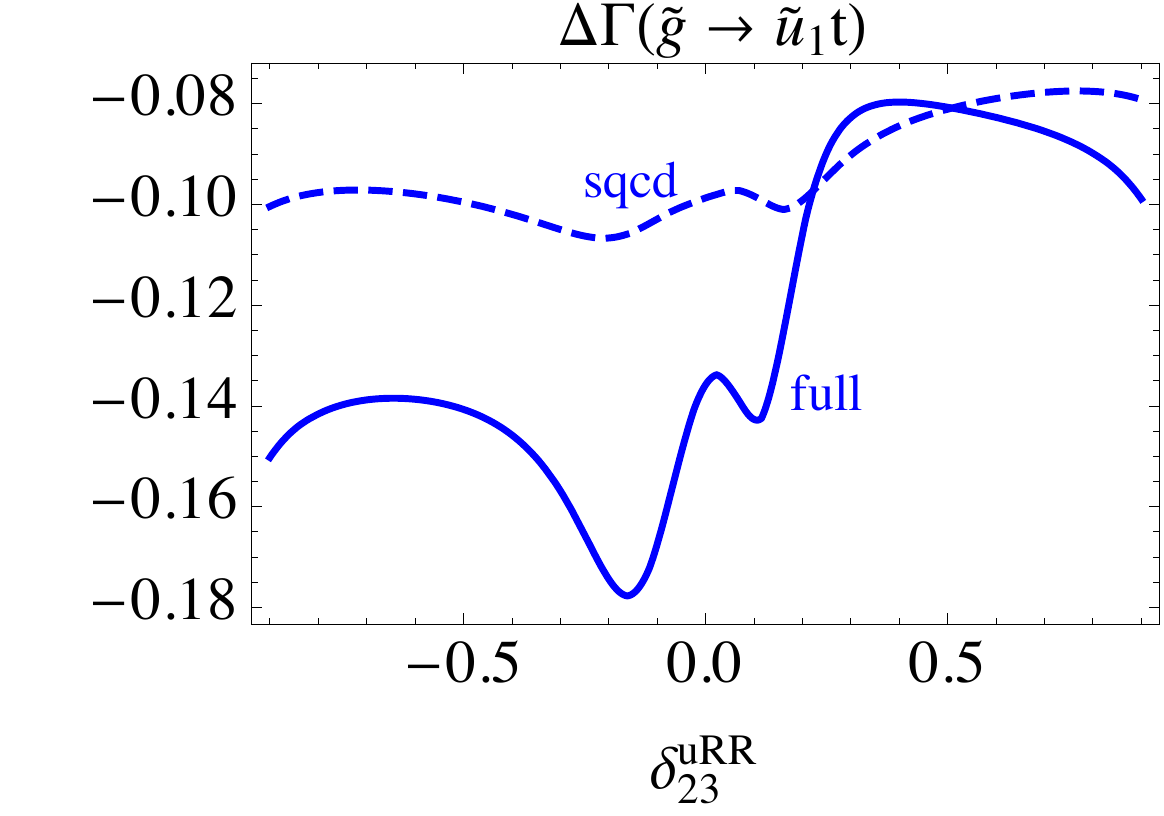}} \hspace*{-0.8cm}}}
   \label{fig1Da}}
\subfigure[]{
   { \mbox{\hspace*{0.5cm} \resizebox{8cm}{!}{\includegraphics{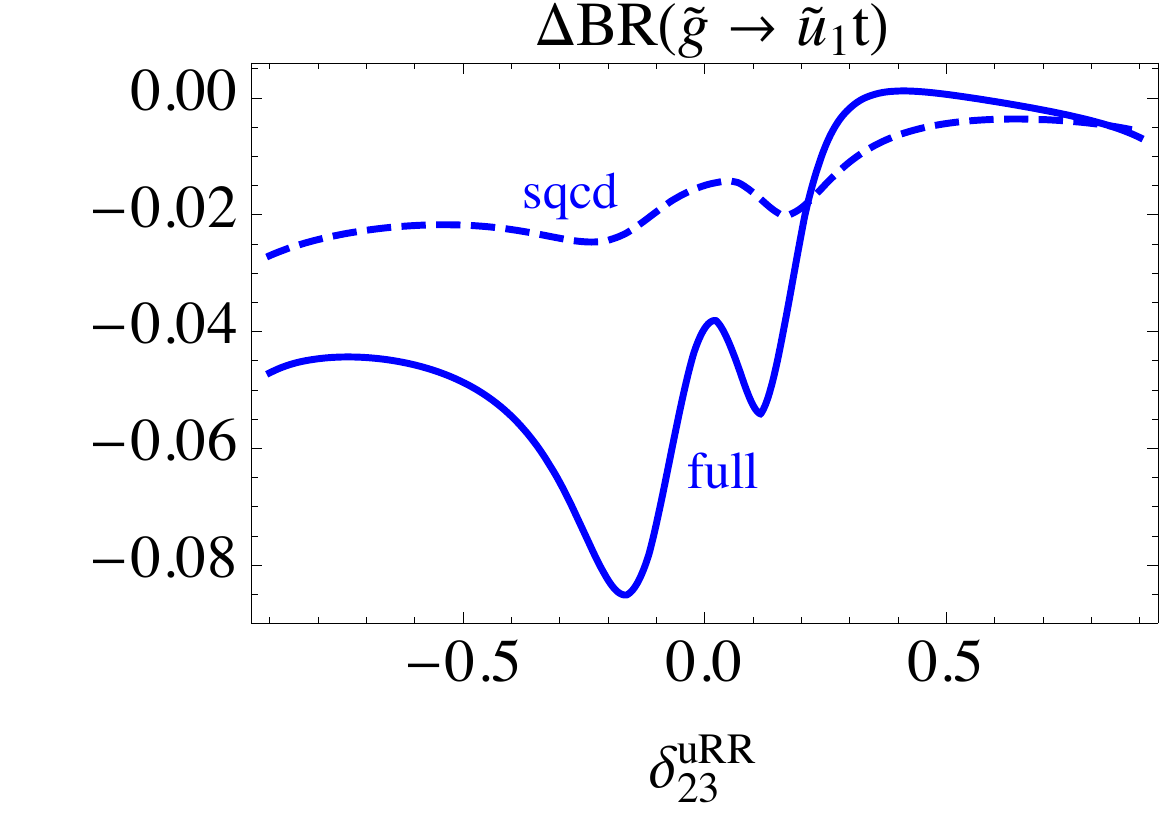}} \hspace*{-0.8cm}}}
   \label{fig1Db}} 
\caption{$\Delta \Gamma$ and $\Delta$BR denote the SQCD
one-loop and the full one-loop corrections relative to
the tree-level result for the decay $\sg \to \su_1 t$
as a function of $\durr$; (a) and (b) is for the partial
width and the branching ratio, respectively.
The other parameters are fixed as in Fig.~\ref{fig1}.}
\label{fig1D}
\end{figure*}

The relative contributions of the one-loop SQCD and the full one-loop part in terms of the tree-level result for 
the decay $\tilde g \to \tilde u_1 t$ as a function of $\durr$ are shown
in Fig.~\ref{fig1Da} for the partial decay width and in Fig.~\ref{fig1Db} for the branching ratio. 
The SQCD corrections  vary in the range of -8\% to - 10\%. 
The EW correction is much stronger dependent on $\durr$, varying between 1\% down to -8\%. 
The effects are similar in the branching ratio (b), but weaker.
Out of the squark masses only $m_{\tilde u_1}$ is strongly dependent on $\durr$. 
The wiggles stem from the complex structures of the QFV one-loop contributions because 
no additional channel opens but those visible in Figs.~\ref{fig1c} and~\ref{fig1d}.

Similar to Figs.~\ref{fig1} and \ref{fig1D}  we have studied the dependence on the sdown right-right parameter
$\ddrr$. The dependence of the total gluino width is similar to that of $\durr$ but the EW contribution is smaller, up 
to 2\%. The individual partial widths show a similar picture with $\su \leftrightarrow \sd$ compared to the ones given 
in Figs.~\ref{fig1c} and  \ref{fig1d}. The corrections to the channels $\sg \to  \sd_1 b$ are smaller than for $\sg \to  \su_1 t$.
It is interesting that for $\sg \to  \su_1 t$ the width varies by about -3\% in the allowed range of $\ddrr$. This effect stems 
from the $\sd$ loops in gluino wave-function correction. 

%
\begin{figure*}[h!]
\centering
   { \mbox{\hspace*{-1cm} \resizebox{8.cm}{!}{\includegraphics{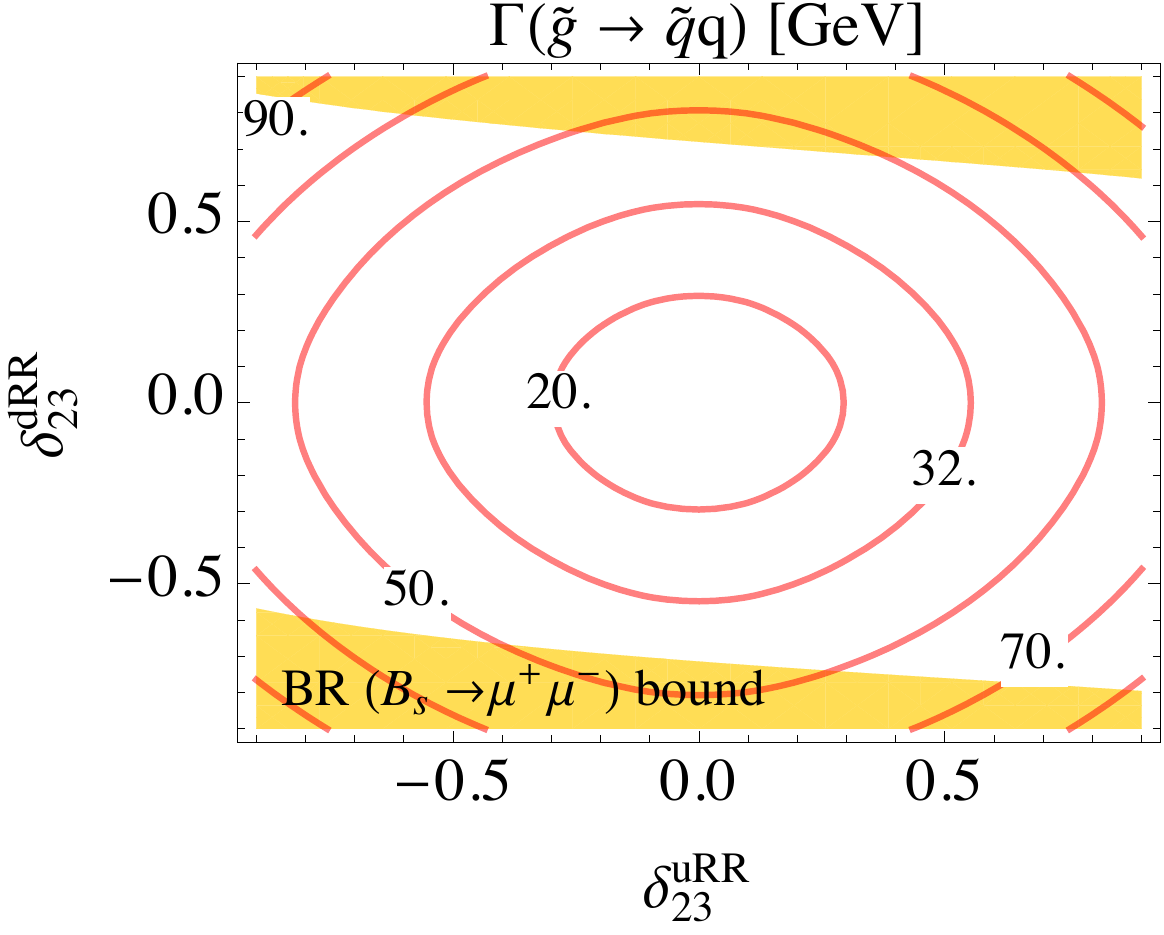}} \hspace*{-0.8cm}}}
\caption{Total two-body decay width $\Gamma(\sg \to \sq q)$ at full one-loop level as a function of the QFV parameters $\ddrr$ and $\durr$.
All the other parameters are given in Table~\ref{basicparam}, except $\durl = \dulr= 0.01$.}
\label{fig3}
\end{figure*}
Fig.~\ref{fig3} shows the dependence of the total gluino width at one-loop level 
on the right-right mixing parameter of both the $\su$ and $\sd$ sectors. The width
varies from 20~GeV up to 90~GeV in the allowed region.

%
\begin{figure*}[ht!]
\centering
\subfigure[]{
   { \mbox{\hspace*{-1cm} \resizebox{8.cm}{!}{\includegraphics{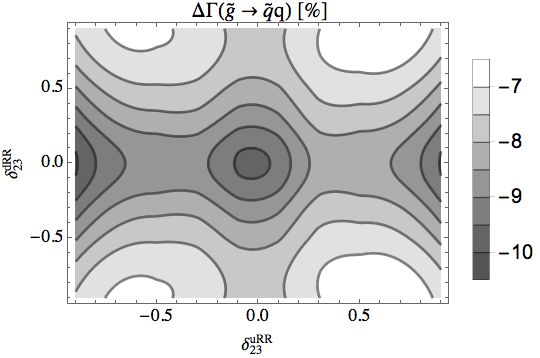}} \hspace*{0.5cm}}}
    \label{fig3Da}}
\subfigure[]{
   { \mbox{\hspace*{-1cm} \resizebox{8.cm}{!}{\includegraphics{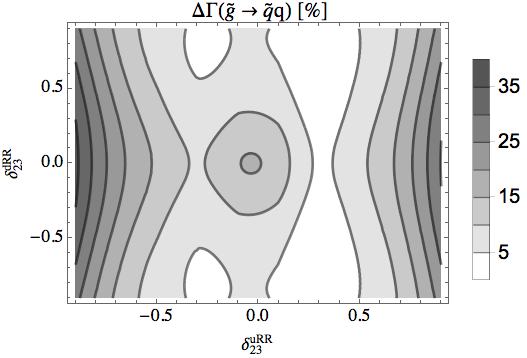}} \hspace*{-0.8cm}}}
    \label{fig3Db}}   
\caption{$\Delta \Gamma$ denotes in (a) the full one-loop contribution in terms of the total tree-level width,
in (b) the EW contribution relative to the SQCD contribution. Both plots are given as a   
function of the QFV parameters $\durr$ and $\ddrr$.
All the other parameters are given in Table~\ref{basicparam}, except $\durl = \dulr= 0.01$.}
\label{fig3D}
\end{figure*}

In Fig.~\ref{fig3Da} the full one-loop part in terms of the tree-level result and in Fig.~\ref{fig3Db} 
the EW contribution relative to the SQCD contribution are shown for the total two-body gluino decay width as a function of $\durr$ and $\ddrr$. 
We see in~\ref{fig3Da} a constant QFC one-loop contribution of $\sim$ -10\% 
and $\sim$~3\% variation due to QFV. 
The EW part can become up to $\sim$~35\% of the SQCD one (\ref{fig3Db})
for large $|\durr|$ where the $\su_1 t$ mode becomes important,
since the $\su_1$ mass becomes smaller due to the $\su$-sector
right-right mixing effect. Furthermore, as $\su_1$ is mainly a top squark,
the EW corrections to the $\su_1 t$ mode are significant, mainly controlled by the large top-quark
Yukawa coupling $Y_t$.

We also have studied the gluino mass dependence.
The full one-loop corrections to the total width are of the order of
about -10\% in the gluino mass range of 2.3 - 4.0 TeV.

\section{Conclusions}
\label{sec:concl}

All gluino two-body decays at full one-loop level in the MSSM with QFV were studied.
Numerically we  have focused on a scenario 
where only the decays to $\tilde u_{1,2}$ and $\tilde d_{1,2}$ are 
kinematically open and $\tilde u_1$ is a mixture of $\tilde c_R$ 
and $\tilde t_R$ controlled by $\delta_{23}^{uRR}$, and $\tilde d_1$ 
is a mixture of $\tilde s_R$ and $\tilde b_R$ controlled by 
$\delta_{23}^{dRR}$. We obey the constraints from B-physics,
the LHC constraints for the masses of the SUSY particles, especially that one for $m_{h^0}$
and have checked also the vacuum stability conditions. 

The full one-loop corrections to the gluino decay widths are mostly 
negative. For the total decay width they are in the range of -10\%  
with a weak dependence on QFV parameters for both SQCD (gluon and gluino) and electroweak (includes also the photon) 
corrections.
For the decay width into $\su_1$ we can have a  total correction
up to -18\%, with the electroweak part up to -8\%, strongly depending on the QFV parameters,
for $\sd_1$ the effects are smaller. In the branching ratios these effects are washed out.
In general, it turns out that the EW corrections can be 
in the range of up to 35\% of the SQCD corrections due to the large top-quark Yukawa coupling.

\section*{Acknowledgments}
This work is supported by the "Fonds zur F\"orderung der
wissenschaftlichen Forschung (FWF)" of Austria, project No. P26338-N27.
We thank S. Frank for providing the
\mbox{program~FVSFOLD}.

\end{document}